\pdfoutput=1
\documentclass[useAMS,usenatbib]{mnras}
\usepackage{anyfontsize}
\usepackage{hyperref}
\usepackage{amsmath}
\usepackage{amssymb}
\usepackage{wasysym}
\usepackage{graphicx}
\usepackage{grffile}
\usepackage{times}
\usepackage[percent]{overpic}
\usepackage{color}
\usepackage{newtxtext}
\usepackage[T1]{fontenc}
\usepackage{ae,aecompl}
\usepackage{pstricks}
\title[Grooves and spiral modes]{Instabilities in disc galaxies:\\from
  noise to grooves to spirals}

\author[S. De Rijcke et al.]{Sven De Rijcke$^{1}$\thanks{E-mail:
    sven.derijcke@Ugent.be}, Jean-Baptiste Fouvry$^{2}$\thanks{Hubble
    fellow, E-mail: fouvry@ias.edu}, Christophe
  Pichon$^{3,4}$\thanks{E-mail: pichon@iap.fr}\\ $^{1}$Ghent University, Dept. Physics \& Astronomy,
  Krijgslaan 281, S9, B-9000, Ghent, Belgium\\ $^{2}$Institute for
  Advanced Study, Einstein Drive, Princeton, NJ 08540,
  USA\\ $^{3}$Institut d'Astrophysique de Paris, 98 bis boulevard
  Arago, F-75014 Paris Cedex, France\\ $^{4}$Korea Institute for
  Advanced Study (KIAS), 85 Hoegiro, Dongdaemun-gu, Seoul, 02455,
  Republic of Korea}

\begin{document}

\def \aj {Astron. J.}
\def \mnras {Mon. Not. R. Astron. Soc.}
\def \apj {Astrophys. J.}
\def \apjs {Astrophys. J. Suppl.}
\def \apjl {Astrophys. J. Let.}
\def \aap {Astron. \& Astrophys.}
\def \aaps {Astron. \& Astrophys. Suppl. Ser.}
\def \nat {Nature}
\def \pasp {Pub. Astron. Soc. Pac.}
\def \araa {ARA\&A}
\date{}
\pagerange{\pageref{firstpage}--\pageref{lastpage}} \pubyear{2009}

\maketitle

\label{firstpage}

\begin{abstract}
Using the linearized Boltzmann equation, we investigate how grooves
carved in the phase space of a half-mass Mestel disc can trigger the
vigorous growth of two-armed spiral eigenmodes. Such grooves result
from the collisional dynamics of a disc subject to finite-$N$ shot
noise, as swing-amplified noise patterns push stars towards
lower-angular momentum orbits at their inner Lindblad
radius. Supplementing the linear theory with analytical arguments, we
show that the dominant spiral mode is a cavity mode with reflections
off the forbidden region around corotation and off the deepest
groove. Other subdominant modes are identified as groove modes. We
provide evidence that the depletion of near-circular orbits, and not
the addition of radial orbits, is the crucial physical ingredient that
causes these new eigenmodes.

Thus, it is possible for an isolated, linearly stable stellar disc to
spontaneously become linearly unstable via the self-induced formation
of phase-space grooves through finite-$N$ dynamics. These results may
help explain the growth and maintenance of spiral patterns in real
disc galaxies.
\end{abstract}

\begin{keywords}
galaxies: kinematics and dynamics -- galaxies: evolution -- galaxies: spiral
\end{keywords}
\newcommand{\beqn}{\begin{equation}}
\newcommand{\neqn}{\end{equation}}
\newcommand{\mathi}{{\mathrm{i}}}
\newcommand{\mathe}{{\mathrm{e}}}

\section{Introduction}\label{intro}

Despite the many tantalizing hints gleaned from detailed observations
\citep{2009ApJ...702..277M,2010ApJ...715L..56S,2018ApJ...854..182B},
analytical calculations
\citep{goly65,juto66,lyn72,1976ApJ...205..363M,omurkanov14,dv16}, and
numerical simulations
\citep{zhang96,2011MNRAS.410.1637S,ovh13,2014ApJ...785..137S,
  2016ApJ...826L..21S,2017ApJ...834....7S}, the cause(s) and the life
expectancy of spiral structures in disc galaxies are still
uncertain. Even restricting the general problem to that of the
gravitational dynamics of an axially symmetric, razor-thin stellar
disc still leaves a wealth of dynamical processes to be explored and
understood. Given the observed variety of the spiral galaxy zoo, with
specimens ranging from the beautifully symmetric grand-design spirals
to the patchy and chaotic flocculent disc galaxies, a unique
explanation may even seem elusive. This is unfortunate, given the
apparent importance of spiral structures for the secular evolution of
the galaxies that host them \citep{zhang98,2002MNRAS.336..785S,
  2008ApJ...684L..79R,2012MNRAS.426.2089R,2014RvMP...86....1S,
  2015A&A...584A.129F,2015MNRAS.450..873V,2015NewA...34...65Z,2018MNRAS.476.1561D}
and for regulating the rate and location of their star formation
\citep{2002MNRAS.337.1113S,2016MNRAS.459.3130A,2017MNRAS.468.1850H}.

Early on in the history of this topic, it was understood that stellar
discs can respond vigorously to disturbances whose wavelength is much
smaller than the host disc. Such localized perturbations can be shown
to be strongly amplified as they shear from leading to trailing
\citep{goly65,juto66,toomre81}. This ``swing amplification'' mechanism
appears to be a plausible explanation for the ragged appearance of the
flocculent disc galaxies. It cannot, at first sight, explain the open,
long-wavelength patterns observed in grand-design systems. In the
original quasi-steady-state theory for grand-design spirals
\citep{linshu64}, the gravitational maintenance of an imposed spiral
pattern was expounded. However, this theory lacked a generating
mechanism for the spiral patterns.

Barring those cases where a spiral pattern can be linked to an
external pertuber's tidal forces, an internal cause must be
searched. It was soon realized that such patterns could originate from
growing eigenmodes in linearly unstable stellar discs
\citep{toomre64,1965MNRAS.129..321H,kalnajs77,b9}. If a galaxy is
linearly unstable, even the smallest perturbation is sufficient to
trigger its eigenmodes
\citep{toomre64,1965MNRAS.129..321H,kalnajs77,1997MNRAS.291..616P,b9}
and begin its evolution towards higher entropy states
\citep{lyn72,zhang96}. This leads to secular radial migration of stars
and disc heating \citep{1999ApJ...518..613Z,2002MNRAS.336..785S,
  2008ApJ...684L..79R,2012MNRAS.426.2089R}.

In this paper, we show that such linearly unstable states need not be
exceptional and that they are, in fact, a natural outcome of the
collisional dynamics of a finite-$N$ stellar disc. This point was
already argued by \citet{2012ApJ...751...44S} and
\citet{2015A&A...584A.129F}, who investigated the evolution of a
linearly stable half-mass Mestel disc
\citep{1963MNRAS.126..553M,toomre81} using $N$-body simulations and
the integration of the inhomogeneous Balescu-Lenard equation
\citep{2010MNRAS.407..355H,2012PhyA..391.3680C}, respectively. In
other words, the finite-$N$ dynamics of an initially linearly stable
stellar disc can lead to its spontaneous destabilization and the
subsequent growth of spiral-shaped eigenmodes. The origin of these
modes lies in the gravitational amplification of waves trapped between
the groove and corotation.

This paper is organized as follows. 
In Section \ref{halfM}, we introduce the half-mass Mestel disc model,
both without and with phase-space grooves. This is followed by our
linear mode analysis of this model in Section \ref{sec:modeMestel}. We
discuss the significance of our results in Section \ref{sec:disc} and
we conclude with Section \ref{sec:conc}. The linear stability tool we
employ is shortly described in Appendix \ref{pystab}.

\section{The half-mass Mestel disc}\label{halfM}

\subsection{The linearly stable half-mass Mestel disc}

The Mestel disc \citep{1963MNRAS.126..553M} has a surface mass density
given by \beqn \Sigma(r) = \Sigma_0 \frac{r_0}{r}, \label{sigma0}
\neqn which self-consistently generates a gravitational field with a
binding potential \beqn V_0(r) = -v_0^2 \ln\left( \frac{r}{r_0}
\right).  \neqn Here, $v_0$ is the constant circular velocity of this
stellar disc model and $r_0$ is a scale length. Clearly, $v_0^2 = 2\pi
G \Sigma_0 r_0$ for consistency (this relation defines the density
scale factor $\Sigma_0$). A self-consistent distribution function, or
DF, for this model exists, of the form \beqn F_{\mathrm{M}}(E,J_\phi) =
\frac{\Sigma_0}{2^{q/2} \sqrt{\pi} r_0^q \sigma^{2+q} \Gamma\left(
  \frac{1+q}{2} \right)} J_\phi^q \exp( E/\sigma^2 ), \neqn with $E$
binding energy, $J_\phi$ angular momentum, and $q$ a real number that
links the radial velocity dispersion $\sigma$ to the circular velocity
$v_0$ via the relation $\sigma^2 = v_0^2/(1+q)$
\citep{1977ARA&A..15..437T}. We adopt the numerical values $q=11.4$
and $r_0=20$, similarly to \citet{2012ApJ...751...44S}.

In order to obtain a stellar disc with a finite total mass and with a
non-diverging central density, we multiply this DF with two cut-out
functions, $H_{\rm inner}(J_\phi)$ and $H_{\rm outer}(J_\phi)$, of the
form
\begin{align}
  H_{\rm inner}(J_\phi) &= \frac{J_\phi^n}{\left(r_{\rm inner} v_0
    \right)^n + J_\phi^n }, \\ H_{\rm outer}(J_\phi) &=
  \frac{\left(r_{\rm outer} v_0 \right)^m}{\left(r_{\rm outer} v_0 \right)^m +
    J_\phi^m}.
\end{align}
Here, we choose $n=4$, $m=5$, $r_{\rm outer}=11.5$, and we adopt units
such that $v_0 = G = r_{\rm inner} = 1$. As in
\citet{2012ApJ...751...44S}, we use a Plummer softening length
$\varepsilon=1/8$ in equation (\ref{plsoft}) for the inter-particle
interaction potential.


The DF is further multiplied with an active fraction $\xi=1/2$, such
that the DF describing the dynamics of the stellar component of the
half-mass Mestel disc has the form \beqn F_0(E,J_\phi) = \xi H_{\rm
  inner}(J_\phi) F_{\mathrm{M}}(E,J_\phi) H_{\rm outer}(J_\phi). \label{F0} \neqn
The rest of the matter, making up the deficit between the stellar
density generated by the DF (\ref{F0}) and the Mestel disc density
(\ref{sigma0}), is supposed to reside in a rigid, unresponsive halo
and bulge. As reported by \citet{toomre81},
\citet{1998MNRAS.300..106E}, \citet{2012ApJ...751...44S}, and
\citet{2015A&A...584A.129F}, this stellar disc is linearly stable, a
fact that we confirmed with our own stability analysis.

\subsection{The grooved half-mass Mestel disc}

\begin{figure}
\includegraphics[trim=5 10 0 0, clip,width=0.49\textwidth]{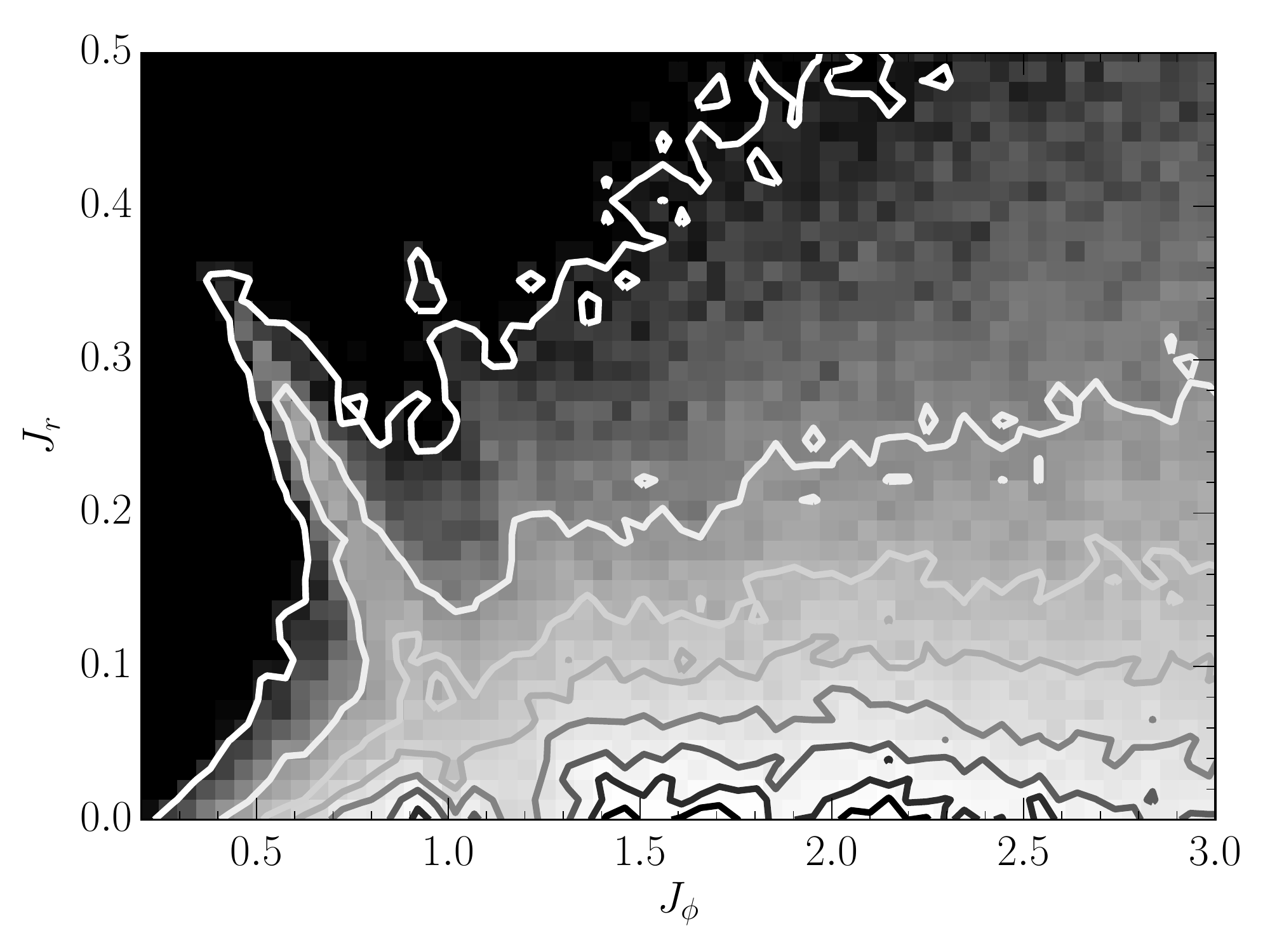}
\caption{The phase-space distribution of the particles in simulation
  50M of \citet{2012ApJ...751...44S} at time 1400, presented as a
  function of angular momentum, denoted by $J_\phi$, and the radial
  action, denoted by $J_r$. Circular orbits are characterized by
  $J_r=0$. The contours of constant stellar phase-space density are
  evenly spaced between 2.5~\% and 95~\% of the maximum value of the
  DF (estimated as the mean of the 10 highest-valued pixels). Three
  grooves can be identified where circular orbits are depopulated
  around the angular momentum values $J_\phi \approx 1.2$, $1.55$, and
  $1.90$.
 \label{fig:Sellwood_sim}}
\end{figure}

While this half-mass Mestel disc is linearly stable, this does not
mean that it cannot evolve secularly through finite-$N$ effects. In
order to maximally elucidate the underlying dynamics,
\citet{2012ApJ...751...44S} and \citet{2015A&A...584A.129F} consider
only gravitational forces up to the $m=2$ harmonic in their studies of
the evolution of this disc model. Therefore, we will likewise only
consider $m=2$ spiral patterns in the remainder.

As shown by \citet{2012ApJ...751...44S} and
\citet{2015A&A...584A.129F}, swing-amplified finite-$N$ stellar
density fluctuations in this Mestel disc transport angular momentum
away from their inner Lindblad resonances (ILR) and scatter stars away
from (near-)circular orbits around certain angular momenta. If a
spiral wave is time-independent in a co-rotating coordinate frame then
each star evolves along a track of constant Jacobi integral, $E_{\mathrm{J}}$,
given by \beqn E_{\mathrm{J}} = E + \Omega_{\mathrm{p}} J_\phi, \neqn with
$\Omega_{\mathrm{p}}$ the wave's pattern speed. The interaction
between the stars and transient swing-amplified patterns induces
stellar migration through action space, changing the stars' binding
energy and angular momentum but preserving their Jacobi integral. At a
pattern's ILR, this pushes stars away from circular orbits towards
orbits with lower angular momentum $J_\phi$ and higher values of the
radial action $J_r$. Hence, the DF develops what we here refer to as
{\em grooves} (loci around constant $E_{\mathrm{J}}$-values where stars have
diffused away from high-$J_\phi$ orbits, leaving behind a depleted DF)
and {\em ridges} (loci around constant $E_{\mathrm{J}}$-values where stars have
diffused towards high-$J_r$ orbits, causing an enhanced DF).

In Figure \ref{fig:Sellwood_sim}, we show the DF of the half-mass
Mestel disc, grooved on the long term by the non-linear evolution of
swing-amplified particle-shot noise, as obtained in simulation 50M of
\citet{2012ApJ...751...44S}\footnote{Based on simulation data kindly
  provided to us by Prof. J. Sellwood.} (see also Figure 7 in
\citet{2012ApJ...751...44S} and Figure 4 in
\citet{2015A&A...584A.129F}). Three grooves can be identified where
circular orbits are depopulated around angular momentum values $J_\phi
\approx 1.2$, $1.55$, and $1.90$.

Concurrent with the appearance of these grooves in action space, the
stellar disc develops a set of vigorously growing spiral patterns
whose amplitudes exponentiate with time (see Figure 2 in
\citet{2012ApJ...751...44S}). The most vigorous among these modes has
a rotation frequency $\omega = m\Omega_{\mathrm{p}} \approx 0.55$ (see
Figure 4 in \citet{2012ApJ...751...44S}). It cannot be an eigenmode of
the ungrooved half-mass Mestel disc because it is known to be linearly
stable. Since the mode already appears in the linear regime it is
unlikely to be caused by non-linear mode coupling
\citep{masset97}. Randomizing the azimuthal particle positions does
not prevent the pattern's growth, quite on the contrary, so any
non-axisymmetric features that grew during the disc's first phase of
evolution cannot have caused this growing mode (see Figure 5 in
\citet{2012ApJ...751...44S}). The fact that it grows exponentially
with time suggests that it is a true eigenmode particular to the
grooved half-mass Mestel disc.

Using linear stability analysis, we now show that the vigorously
exponentiating patterns observed by \citet{2012ApJ...751...44S} are
indeed eigenmodes of the grooved half-mass Mestel disc. For this, we
use {\sc pyStab}, a {\sc Python}/{\sc C++} computer code to analyse
the stability of a responsive, self-gravitating razor-thin stellar
disc embedded in the gravitational field of a rigid axisymmetric or
spherically symmetric central bulge and dark-matter halo. The details
of the mathematical formalism behind this code and of its
implementation can be found in \citet{b9}, \citet{dury08}, and
\citet{dv16} so we will not repeat these here in detail. We provide a
brief overview of the formalism and of the employed values of some
numerical parameters of the code in Appendix \ref{pystab}.

\section{Mode-analysis of the grooved half-mass Mestel disc} \label{sec:modeMestel}

\subsection{The fiducial grooved half-mass Mestel disc}

We first mimicked the grooves visible in the DF of simulation 50M (see
Fig. \ref{fig:Sellwood_sim}) of \citet{2012ApJ...751...44S} by
multiplying the DF $F_0(E,J_\phi)$ with ``groove functions'' of the
form \beqn f_{\rm groove}(x) = 1 + \frac{a }{(1-x)^\alpha} + b + cx +
dx^2,\label{fgroove} \neqn with $x$ varying along a line of constant
Jacobi integral, normalized such that $x=0$ at the circular orbit and
$x=1$ at the radial orbit. Each groove function differs from 1 only
inside a narrow region of width $2w$ centered on a given Jacobi
integral. If $f_{\rm groove}(x) > 1$, the phase-space density is
increased; if $f_{\rm groove}(x) < 1$, the phase-space density is
decreased.

Based on Figure \ref{fig:Sellwood_sim}, we identify three angular
momenta where circular orbits are significantly depopulated:~at
$J_\phi \approx 1.2$ (first groove), $1.55$ (second groove), and $1.9$
(third groove), corresponding to the ILR of waves with pattern speeds
$\Omega_{\mathrm{p}} = 0.244$, $0.189$, and $0.154$, respectively. We will
henceforth refer to the half-mass Mestel disc with these three grooves
as the fiducial grooved half-mass Mestel disc.

\begin{table}
	\centering
	\caption{The parameter values employed in the ``groove
          function'' $f_{\rm groove}$ that defines the profile of each
          groove (cf. equation \ref{fgroove}).}
	\label{tab:groove_table}
\begin{tabular}{l|c|c|c|c|c|c}\hline
  & $\alpha$ & $w$ & $a$ & $b$ & $c$ & $d$  \\ \hline
  first groove  & 11 & 0.29 & 0.474 & -0.044 & -5.219 & -44.796 \\
  second groove &  6 & 0.10 & 0.562 &  0.388 & -3.371 & -15.619 \\
  third groove  & 6 &  0.12 & 0.081 &  0.869 & -0.486 & -3.242 \\  \cline{1-7}
\end{tabular}
\end{table}

\begin{figure}
\includegraphics[trim=15 15 0 0, clip,
  width=0.48\textwidth]{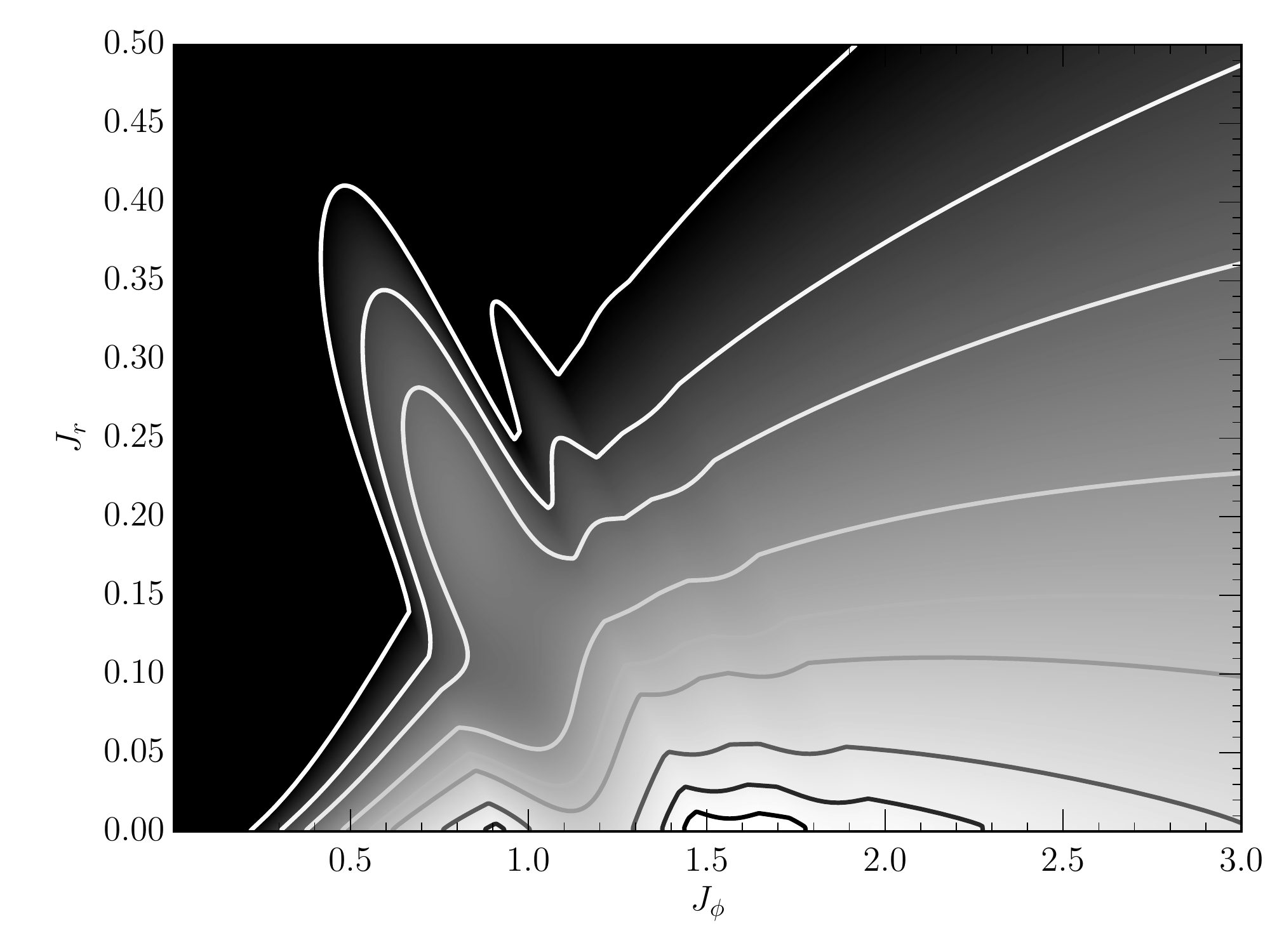}
\caption{The DF of the fiducial grooved half-mass Mestel disc.  The
  contours of constant stellar phase-space density are evenly spaced
  between 2.5~\% and 95~\% of the maximum value of the DF. 
 \label{fig:3grooves}}
\end{figure}

At the first groove, at $J_\phi \approx 1.2$, the value of the DF
decreases by 57~\% while, at the second and third groove, at $J_\phi
\approx 1.55$ and $1.9$, the value of the DF decreases by 5~\%. The
first groove leads to a marked population increase of the lower
angular momentum/higher radial action orbits, which is mimicked here
by setting $\alpha=11$ in equation (\ref{fgroove}). The second and,
especially, the third groove show much less pronounced DF increases
towards low angular momentum orbits so we choose $\alpha=6$ for
these. For each groove, the parameters $w$, $a$, $b$, $c$, and $d$ are
chosen such that the grooves visible in Figure \ref{fig:Sellwood_sim}
are adequately reproduced, cf.  Figure \ref{fig:3grooves} and Table
\ref{tab:groove_table}. We took care to conserve the total mass of the
stellar disc to within a few tenths of a percent. 
\begin{table}
	\centering
	\caption{The eigenmodes of the half-mass Mestel disc with
          three grooves at $J_\phi=1.2$, $1.55$, and $1.9$. The
          complex mode frequency is denoted by $\omega$. The radii of
          the main resonances (ILR, CR, OLR) are indicated and
          resonances that (approximately) overlap with the position of
          a groove are printed in boldface. The physical nature of
          each mode (global or groove mode) is given in the last column. }
	\label{tab:mode_table}
\begin{tabular}{c|c|c|c|c}\hline
  $\omega$ & ILR & CR & OLR & type \\ \hline
  $0.597 + 0.013 {\mathi}$ & 0.98 & 3.35 & 5.72 & global mode  \\
  $1.662 + 0.004 {\mathi}$ & 0.35 & {\bf 1.20} & 2.01  & groove mode\\
  $0.465 + 0.001 {\mathi}$ & {\bf 1.26} & 4.30 & 7.34  &global mode \\
  $1.101 + 0.000 {\mathi}$ & 0.53 & {\bf 1.82 }&  3.10 & groove mode\\
  \cline{1-5}
\end{tabular}
\end{table}

\subsection{Linear stability analysis of the fiducial grooved Mestel disc}

\begin{figure*}
\includegraphics[trim=125 5 15 0, clip,
  width=0.98\textwidth]{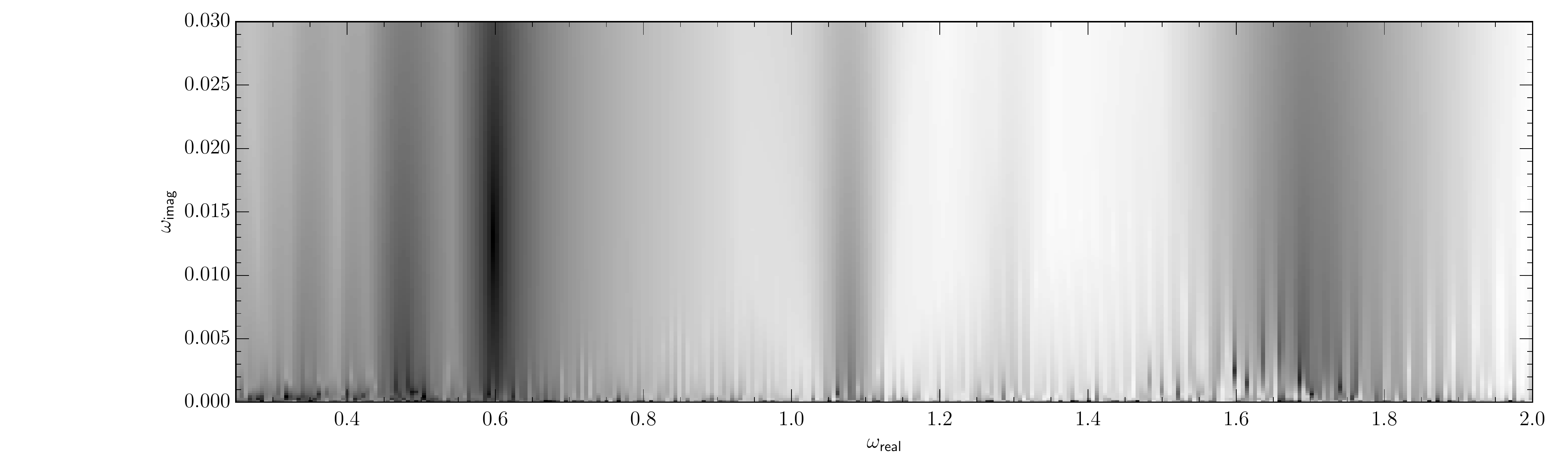}
\caption{The quantity $\min \left| \lambda(\omega)-1 \right|$ in the
  complex frequency plane of the fiducial grooved half-mass Mestel
  disc. An eigenmode lives wherever this quantity is zero (dark
  regions).
 \label{fig:freqplane}}
\end{figure*}

As explained in Appendix \ref{pystab}, each eigenmode corresponds to a
complex frequency $\omega$ for which the matrix ${\mathcal
  C}(\omega)$, defined in equation (\ref{Cdef}), has a unity
eigenvalue. In Figure \ref{fig:freqplane}, we plot the quantity $\min
\left| \lambda(\omega)-1 \right|$ in the complex frequency plane of
the fiducial grooved half-mass Mestel disc. This quantity is zero at
the loci of the eigenmodes.

The complex frequencies $\omega$ of the dominant eigenmodes retrieved
with {\sc pyStab} for the fiducial grooved half-mass Mestel disc are
listed in Table \ref{tab:mode_table}, along with the radii where the
main resonances (ILR, CR, OLR) occur, with CR the corotation resonance
radius and OLR the Outer Lindblad Resonance radius. Resonances that
(approximately) overlap with the position of a groove are printed in
boldface. For the Mestel disc, the resonance radii
$r_{\mathrm{res},\ell}$ of an $m$-armed mode with pattern speed $m
\Omega_{\mathrm{p}} = \omega_{\mathrm{real}}$, with
$\omega_{\mathrm{real}}$ the real part of the mode frequency, are
given by
\begin{equation}
  r_{\mathrm{res},\ell} = \left( m + \sqrt{2}\ell \right)
  \frac{1}{\omega_{\mathrm{real}}}.
\end{equation}
Here, $\ell=0$ gives the CR while $\ell = \pm 1$ give the Lindblad
resonances.


\begin{figure}
\includegraphics[trim=40 5 90 12, clip, width=0.46\textwidth]{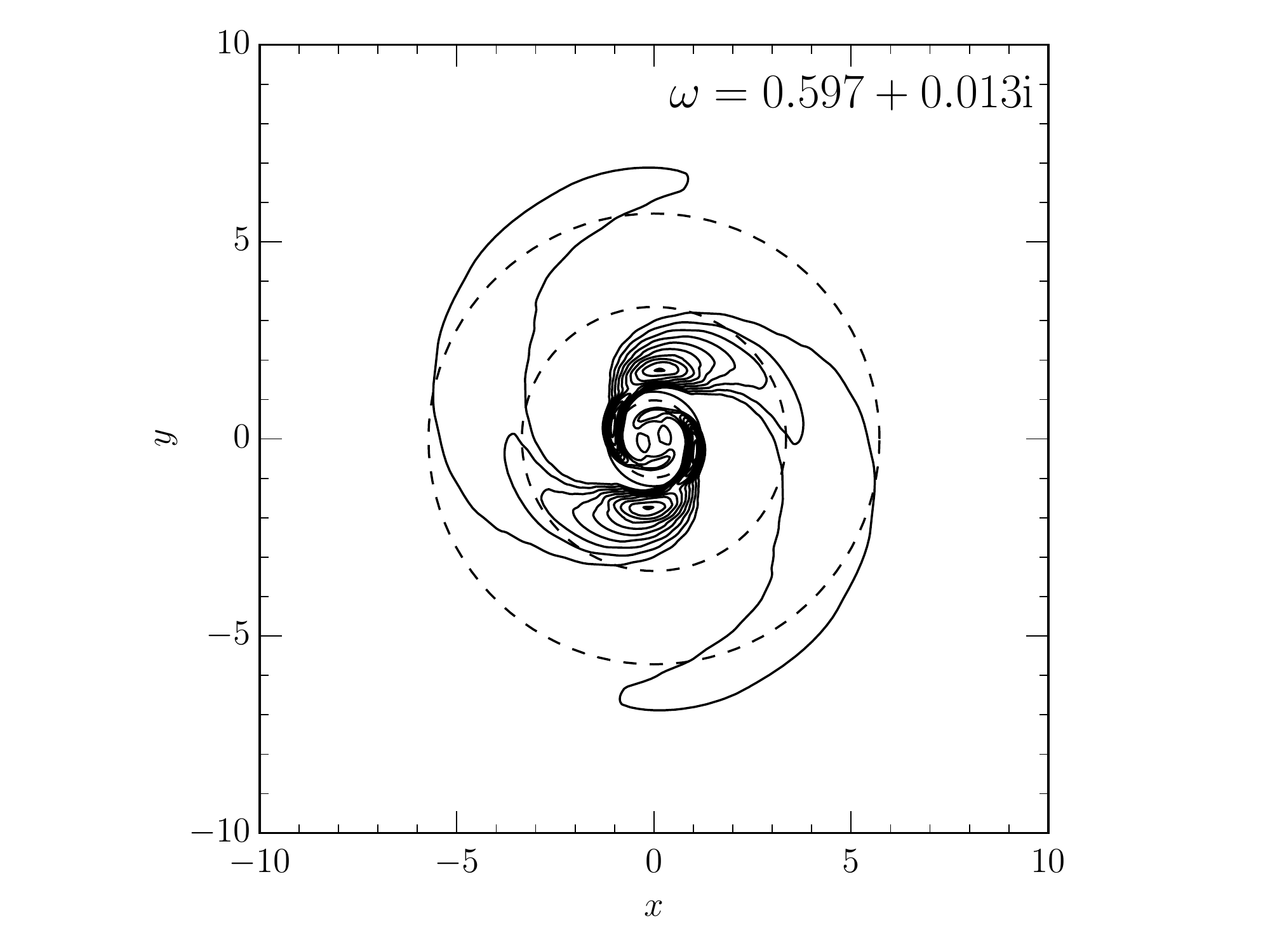}
\put(-230,-210){\includegraphics[trim=-85 -14 0 0, clip, width=0.46\textwidth]{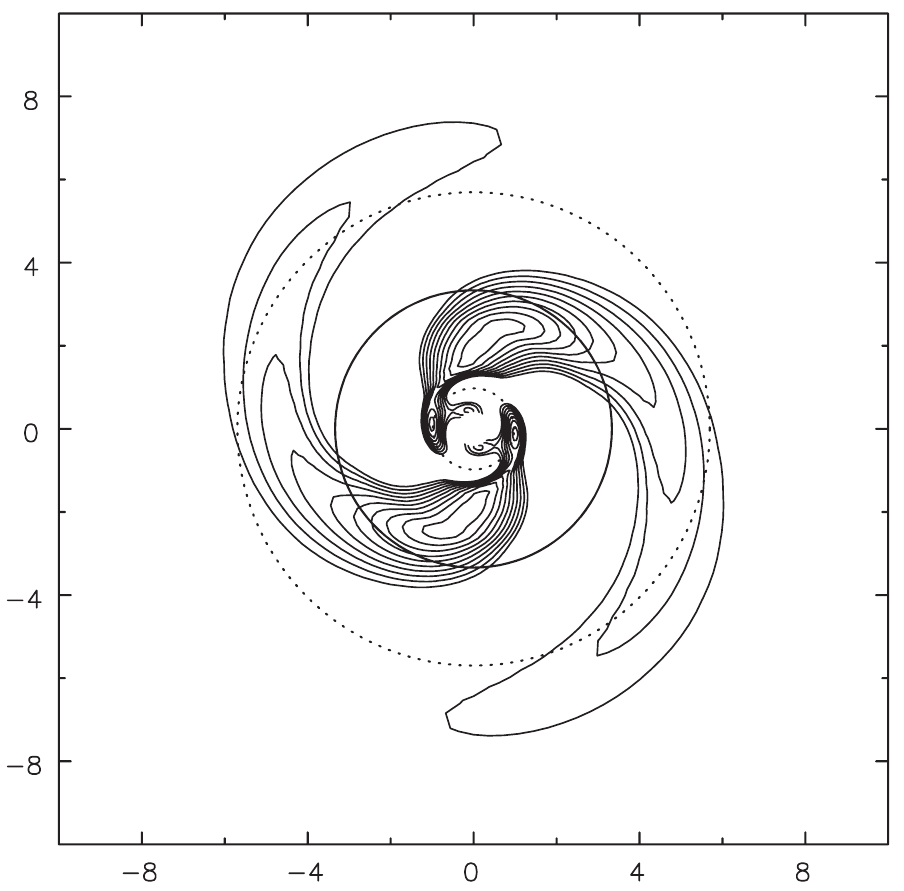}}
\put(-95,-22.5){\fontsize{11.5}{16}$\omega=0.599+0.016{\mathi}$}
\caption{Top panel:~the positive part of the surface density
  perturbation of the $m=2$ global mode with frequency $0.597 + 0.013
  {\mathi}$ of the fiducial grooved half-mass Mestel disc. The contour
  values increase linearly between 1~\% and 99~\% of the maximum
  surface density value. The full line indicates the position of the
  first, main groove and dashed lines mark the mode's ILR, CR, and
  OLR. Bottom panel:~the surface density perturbation of mode 50Mc
  from the $N$-body simulations by \citet{2012ApJ...751...44S}. Dotted
  lines mark the Lindblad resonances while the full line marks the CR
  radius.
 \label{fig:mainmode}}
\end{figure}

\subsection{The $\omega=0.597 + 0.013 {\mathi}$ mode:~a cavity mode}

The most rapidly growing global spiral mode has a pattern speed
$m\Omega_{\mathrm{p}}=0.597$ which places its ILR at a radius of 0.98,
very near the inner edge of the first groove, while its OLR occurs at
a radius of 5.72. The other global spiral mode, with a pattern speed
$m\Omega_{\mathrm{p}}=0.465$, has a negligible growth rate (this is
actually the fastest growing member of a cluster of eigenmodes with
pattern speeds between 0.45 and 0.5, cf. Figure
\ref{fig:freqplane}). The surface density of the dominant global
spiral mode is plotted in Figure \ref{fig:mainmode} (top panel) where
we compare it with the dominant spiral pattern found in the $N$-body
simulations reported in \citet{2012ApJ...751...44S} (bottom
panel). The frequency of the emerging pattern is
estimated\footnote{Private communication with J. Sellwood.} at $\omega
\approx 0.577 + 0.018 {\mathi}$ in simulation 50M and at $\omega
\approx 0.599 + 0.016 {\mathi}$ in simulation 50Mc. We find $\omega =
0.597 + 0.013 {\mathi}$ for the dominant eigenmode of the fiducial
grooved Mestel disc. This good agreement between the pattern speeds
leads us to confirm that we are, in fact, seeing the same mode in the
linear stability analysis and in the $N$-body simulations.

\subsubsection{Mode generating mechanism:~linear stability analysis}

As noted by \citet{2014ApJ...785..137S}, a phase-space groove induces
a spatially localized change of the velocity with which spiral waves
can be radially propagated through the disc. A groove locally changes
the ``impedance'' of the stellar disc. Using the suggestive analogy of
waves being partially reflected at the junction between two strings
with different mass densities (and hence different wave velocities),
these authors argue that stellar density waves impacting radially
inward on a phase-space groove will, likewise, be partially reflected
outward again. This creates the possibility of setting up a resonant
cavity within which inwards moving waves are (partially) reflected
back out again by the groove and outwards moving waves are reflected
back inwards at corotation, thus creating a standing wave pattern.

In the WASER feedback cycle
\citep{1969ApJ...158..899T,1976ApJ...205..363M}, an ingoing
short-wavelength trailing wave is reflected from the unresponsive,
dynamically hot disc interior (if the Toomre $Q$ parameter
\citep{toomre64} rises sufficiently rapidly towards small radii) into
an outgoing long-wavelength trailing wave. At CR, this wave
overreflects into an ingoing and an outgoing short-wavelength traling
wave. Hence, the WASER only involves trailing waves. Moreover, it
leads to rather modest wave amplifications and then only for low $Q$
values. Much more impressive amplifications can be obtained by the
swing amplifier feedback cycle
\citep{1977ARA&A..15..437T,toomre81,1989ApJ...338..104B}. In this
theory, an ingoing trailing wave is reflected from the galaxy center
as an outgoing leading wave. Around CR, this wave is overreflected
into an outgoing and an ingoing trailing wave at which point
amplification factors of $1-2$ orders of magnitude can be
achieved. Significant amplification occurs for discs with $Q \apprle
2$. Interference between the leading and trailing waves within the
resonant cavity is expected to produce a growing eigenmode with a
``lumpy'' density distribution \citep{1984PhR...114..319A,bintrem}.

If the inner $Q$-profile of the half-mass Mestel disc were steep
enough to reflect ingoing waves outwards again, it would already have
done so in the ungrooved model and that model is linearly
stable. Therefore, it is unlikely that the WASER feedback cycle is
responsible for the growth of the global modes in the grooved
half-mass Mestel disc. On the other hand, the Toomre $Q$ parameter is
smaller than $2.0$ in a large part of disc, between radii in the range
$\approx 1.4-9.0$, and hovers around a value of $1.5$ between radii in
the range $\approx 2.0-6.0$. From the famous ``dust to ashes'' figure
of \citet{toomre81}, it is clear that swing amplification in the
$Q=1.5$, $X=k_{\rm crit}r/m=2$ half-mass Mestel disc without inner or
outer cut-outs can boost wave amplitudes by a factor $\sim 30$ (which
significantly hastens the disc's secular dynamics). Unfortunately,
those amplified waves are subsequently absorbed at the ILR. This is,
however, not an issue for the {\em grooved} half-mass Mestel disc. In
the case of waves with a pattern speed $m\Omega_{\mathrm{p}} = 0.597$,
corresponding to the dominant global mode of the fiducial grooved
half-mass Mestel disc, the ILR (at a radius of 0.98) is shielded by
the reflective first groove (around a radius of 1.2).  Moreover, this
dominant global mode clearly has a ``lumpy'' density distribution
(cf. Figure \ref{fig:mainmode}), indicative of interference between
leading and trailing waves.

\begin{figure}
\includegraphics[trim=40 5 90 12, clip, width=0.46\textwidth]{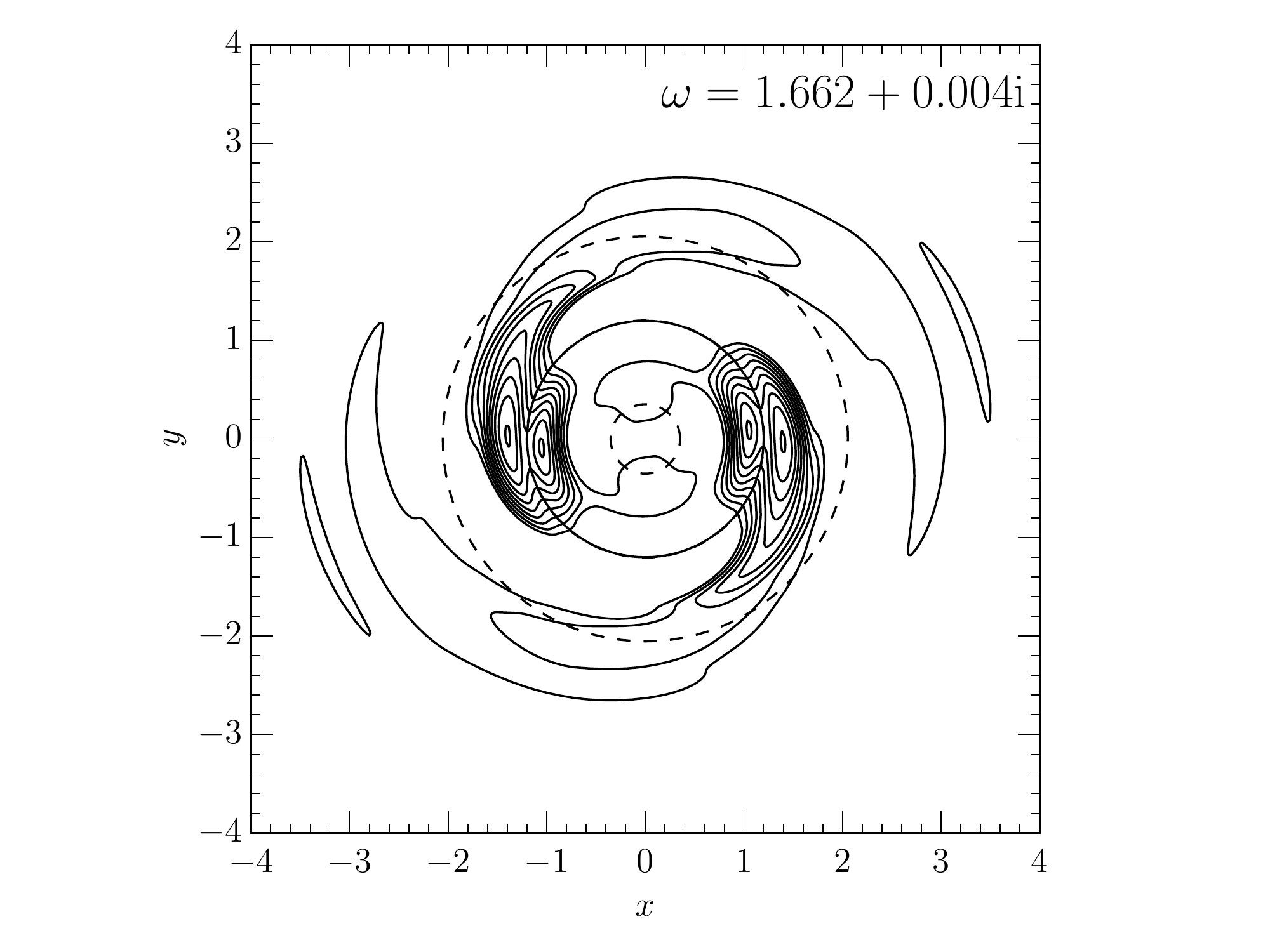}
\caption{The positive part of the surface density perturbation of the
  $m=2$ mode of the fiducial grooved half-mass Mestel disc, with
  frequency $\omega = 1.662 + 0.004{\mathi}$. The contour values
  increase linearly between 1~\% and 99~\% of the maximum surface
  density value. The dashed lines mark the modes' ILR, CR, and OLR and
  the full line indicates the position of the first, main groove.
 \label{fig:3othermodes}}
\end{figure}

\subsubsection{Mode generating mechanism:~WKB arguments} \label{anarg}

If an eigenmode is indeed a standing wave built from wave packets
traveling inside a resonance cavity between radii $r_\mathrm{inf}$
and $r_\mathrm{sup}$ then it must obey the quantum condition
\begin{equation}
  \oint_{r_\mathrm{inf}}^{r_\mathrm{sup}} k\left(\omega_{\mathrm{real}},r\right) \mathrm{d}r = \pi \label{quantum}
\end{equation}
where $k$ is the radial wavenumber of the waves. The mode's growth
rate $\omega_{\mathrm{im}} = \Im\{\omega\}$ then follows from the
relation
\begin{equation}
\frac{1}{\omega_{\mathrm{im}}} \oint_{r_{\mathrm{inf}}}^{r_{\mathrm{sup}}} \frac{\mathrm{d}r}{v_{\mathrm{g}}(r)} = \frac{\ln 2}{2}, \label{wi}
\end{equation}
with $v_{\mathrm{g}}(r)$ the radial group velocity of the traveling
wave patterns \citep{mark77,2014dyga.book}. Hence, the narrower the
cavity and the larger the group velocity, the stronger the
growth. These integrals are to be performed over a full circle between
their lower and upper bounds.

In the WKBJ limit, the dispersion relation for traveling waves is
given by
\begin{equation}
  \kappa^2\left( 1-s^2\right) = 2 \pi G \Sigma \left| k \right| {\mathcal F}(s,\chi) \mathrm{e}^{-\varepsilon\left| k \right|}. \label{drel}
\end{equation}
Here,
\begin{equation}
  s = \frac{\omega_{\mathrm{real}}-m\Omega}{\kappa} \,\, \mathrm{and} \,\,
  \chi = \left( \frac{k\sigma_r}{\kappa} \right)^2,
\end{equation}
with $\Omega$ and $\kappa$ the angular and epicycle frequency of
near-circular orbits, respectively
\citep{1965PhDT.........1K,1966PNAS...55..229L}. The reduction factor
on the right-hand side of this equation is the product of two
factors:~an exponential factor caused by the Plummer softening with
softening length $\varepsilon$ \citep{1994A&A...286..799R}, for which
we adopt the value $\varepsilon=1/8$, and the well-known reduction
factor
\begin{equation}
  {\mathcal F}(s,\chi) = 2\left(1-s^2\right) \frac{\mathrm{e}^{-\chi}}{\chi} \sum_{n \ge 1} \frac{I_n(\chi)}{1-\frac{s^2}{n^2}},
\end{equation}
with $I_n$ a modified Bessel function of the first kind, due to the
stellar velocity velocity dispersion. This latter expression is only
correct for a Schwarzschild DF so the following results must be
regarded as approximate. The group velocity $v_{\mathrm{g}}$ of a wave packet
centered on wave number $k$ is given by
\begin{equation}
  v_{\mathrm{g}} = \frac{\partial \omega_{\mathrm{real}}}{\partial k}.
\end{equation}
From the dispersion relation stated above, we can derive the relation
\begin{align}
  \left[ \frac{2s}{1-s^2} + \mathrm{e}^{\varepsilon \left| k \right|} \frac{\partial \ln \mathcal{F}}{\partial s}
    \right] v_{\mathrm{g}} = \nonumber \\
  & \hspace*{-4em}-\frac{\kappa}{k} \left[ 1- \varepsilon \left| k \right| + 2 
    \mathrm{e}^{\varepsilon \left| k \right|} \frac{\partial \ln \mathcal{F}}{\partial \ln \chi} \right]
  \label{vgroup}
\end{align}
for the group velocity. The partial derivatives of the reduction
factor $\mathcal{F}$ that figure in this equation can be evaluated
analytically as
\begin{align}
  \frac{\partial \mathcal{F}}{\partial \chi} &= \left(1-s^2\right)
  \frac{\mathrm{e}^{-\chi}}{\chi} \sum_{n \ge 1}
  \frac{I_{n-1} -2\left[1+\frac{1}{\chi}\right]I_n + I_{n+1} }{1-\frac{s^2}{n^2}}, \nonumber \\
  \frac{\partial \mathcal{F}}{\partial s} &= -4s\frac{\mathrm{e}^{-\chi}}{\chi} \sum_{n \ge 2}
  \frac{ 1-\frac{1}{n^2}}{1-\frac{s^2}{n^2}} I_n,
\end{align}
where all Bessel functions have $\chi$ as argument. We terminate the
infinite sums in the evaluation of this reduction factor and its
derivatives when a relative accuracy of $10^{-6}$ has been obtained.

We now wish to compute the frequency of a growing standing wave in the
half-mass Mestel disc with a groove at a radius $r_{\mathrm{groove}}$,
which is supposed to reflect incoming wave packets back out again,
using this analytical theory. For a given real frequency
$\omega_{\mathrm{real}}$, we first compute the inner radius of the
forbidden region around CR, where the dispersion relation (\ref{drel})
admits no real solutions for $k$. This we equate to the outer radius
of the resonance cavity, $r_\mathrm{sup}$. For the inner radius,
$r_\mathrm{inf}$, we use the maximum of the groove radius and the
wave's ILR radius. In practice, this always turned out to be the
groove radius.

For any radius $r$ inside the resonance cavity, we then find the two
positive solutions $k_{\mathrm{short}}(\omega_{\mathrm{real}},r)$ and
$k_{\mathrm{long}}(\omega_{\mathrm{real}},r)$ of the dispersion
relation (\ref{drel}). The quantum condition (\ref{quantum}) can be
rewritten as
\begin{align}
  \oint_{r_\mathrm{inf}}^{r_\mathrm{sup}}
  k\left(\omega_{\mathrm{real}},r\right)\mathrm{d}r & \nonumber
  \\ &\hspace*{-8em} = \int_{r_\mathrm{inf}}^{r_\mathrm{sup}} \left[
    k_{\mathrm{short}}(\omega_{\mathrm{real}},r)+k_{\mathrm{long}}(\omega_{\mathrm{real}},r)\right]\mathrm{d}r
  =\pi.
  \end{align}
Numerical root finding finally allows us to solve this equation for
the real frequency $\omega_{\mathrm{real}}$ in case the disc supports
a cavity mode. Equation (\ref{vgroup}) yields the group velocities,
$v_{\mathrm{g},{\mathrm{short}}}(r)$ and $v_{\mathrm{g},{\mathrm{long}}}(r)$, of the
short and long branch waves, respectively. These are then inserted
into equation (\ref{wi}), which can be rewritten as
\begin{align}
  \oint_{r_{\mathrm{inf}}}^{r_{\mathrm{sup}}}
  \frac{\mathrm{d}r}{v_{\mathrm{g}}(r)} 
  &=\int_{r_{\mathrm{inf}}}^{r_{\mathrm{sup}}}
  \left[
    \frac{1}{v_{\mathrm{g},{\mathrm{short}}}(r)}+\frac{1}{v_{\mathrm{g},{\mathrm{long}}}(r)}
    \right] \mathrm{d}r \nonumber \\ & = \frac{\ln
    2}{2}\omega_{\mathrm{im}}. \label{wi2}
\end{align}
This, finally, provides us with an estimate for the growth rate
$\omega_{\mathrm{im}}$ of the growing mode.

For the fiducial value of the groove radius,
$r_{\mathrm{groove}}=1.2$, this yields the analytical frequency
estimate $\omega = 0.510 + 0.017\mathrm{i}$. This is tantalizingly
close to the value $\omega = 0.597 + 0.013\mathrm{i}$ that we derived
from the full linear theory. Together with all the arguments given
above, this can be taken as evidence for the idea that the dominant
global mode of the grooved Mestel disc is caused by swing-amplified
traveling wave packets inside a resonance cavity between the groove --
rendered reflective by the depopulation of near-circular orbits -- and
the mode's CR.


\subsection{The $\omega=1.662 + 0.004 {\mathi}$ mode:~a groove mode}

Glancing back at Table \ref{tab:mode_table}, the second most unstable
mode of the grooved Mestel disc, depicted in Figure
\ref{fig:3othermodes}, has properties that suggest that it is a groove
mode caused by the first groove\footnote{Private communication with
  J. Sellwood:~power spectra of the simulations described in
  \citet{2012ApJ...751...44S} show the existence of a slowly growing
  pattern with pattern speed $m\Omega_{\mathrm{p}} \approx 1.65$ when
  extended to higher pattern frequencies than was reported in Figure 4
  of that work.}, depopulating circular orbits around angular
momentum $J_\phi=1.2$. There actually appears to exist a cluster of
eigenmodes with pattern speeds very close to this value, cf. Figure
\ref{fig:freqplane}, of which this is the most rapidly growing member.

\citet{Sellwood91} describe growing instabilities with their CR at a
sharp groove, i.e. a local depression in the stellar density
distribution, that are an instance of the negative-mass instability
\citep{1978ApJ...221...51L}. These authors show that such groove modes
originate from the coupling of two waves, one on each side of the
groove, with their corotation resonance inside the groove. The wavy
displacements of disc material at the groove's edges couple across the
groove through gravity. Thus, local overdensities can exchange angular
momentum in such a way that material on the groove's inner edge gains
angular momentum and is pulled outwards while material on the groove's
outer edge loses angular momentum and is pulled inwards. This
counteracts the disc's shear and facilitates mass clumping. Thus, the
amplitude of these mass displacements grows exponentially with
time. The rest of the pattern can then be explained as the disc's
response to these growing overdensities via the mechanism described by
\citet{juto66}.

\begin{figure}
\includegraphics[trim=40 15 45 0, clip, width=0.5\textwidth]{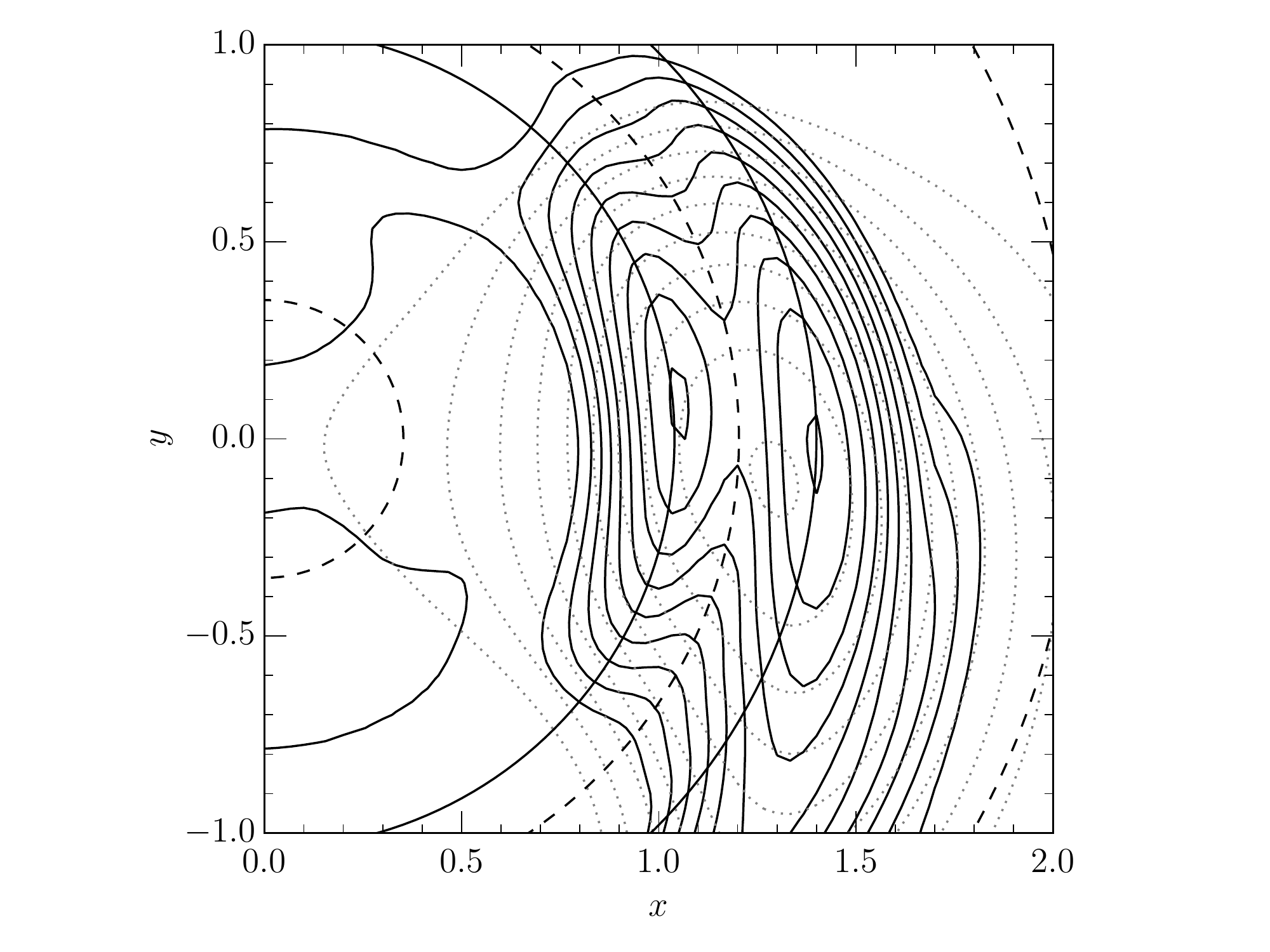}
\caption{Close-up view of the density peak of the $\omega = 1.662 +
  0.004{\mathi}$ groove mode from Fig. \ref{fig:3othermodes}. Full-line contours trace the surface
  density perturbation (in 10 equidistant steps between 1~\% and 99~\%
  of its maximum value) while the dotted contours trace the mode's
  gravitational binding potential (in 10 equidistant steps between
  1~\% and 99~\% of its maximum value). Dashed circles mark the loci
  of the mode's ILR, CR, and OLR. The two full-line circles demarcate
  the region where the DF has been reduced by more than 50~\%, which
  could be taken as a measure for the width of the groove.
 \label{fig:angmom_1stgroove}}
\end{figure}

The CR of the $\omega=1.662 + 0.004 {\mathi}$ mode indeed sits
squarely inside the first groove, cf. Figure
\ref{fig:angmom_1stgroove}. It is also clear from this figure that the
mode, as expected for a groove mode, consists of two density peaks
sitting on either side of the groove with a gravitational potential
peak between them. This suggests this is a groove mode caused by the
first groove. There also appears to exist a neutral eigenmode with
pattern speed $m\Omega_{\mathrm{p}}=1.101$ which can, likewise, be
interpreted as a groove mode caused by the third groove. Its vanishing
growth rate explains why it did not appear in the simulations reported
by \citet{2012ApJ...751...44S}.

\section{Discussion} \label{sec:disc}

\subsection{The number of grooves}

The presence of the minor second and third groove has little bearing
on the existence of the modes although it does slightly impact their
frequencies. For instance, if we remove these two minor grooves,
keeping only the deep and wide first groove, the main global mode is
retrieved with a frequency $\omega=0.603 + 0.016 {\mathi}$, with its
ILR sitting at a radius of 0.97, still just inside the inner edge of
the groove. The groove mode now has a frequency $\omega=1.695 + 0.003
{\mathi}$. Its CR, now at a radius 1.18, again falls inside the
groove. The remaining groove is obviously instrumental to the
existence of these modes. Without it, the grooveless half-mass Mestel
disc is, as already mentioned, linearly stable.

\subsection{The groove profile}

A groove consists both of a depletion of circular orbits and an
overpopulation of more radial orbits. In order to test which of these
two features is actually causing the modes in the grooved Mestel disc,
we first computed the eigenmode spectrum of a model with the fiducial
groove function modified according to the prescription \beqn f_{\rm
  groove}(x) \rightarrow \max \left\{ 1, f_{\rm groove}(x) \right\}
\neqn so that no stars are ever removed but only added. This removes
the grooves at the loci of near-circular orbits but leaves the ridges
at more radial orbits in place. This particular model turned out to
have a single (barely) growing global eigenmode at $\omega=0.611 +
0.001 {\mathi}$. The high-frequency mode of the fiducial model is
completely absent.

Next, we computed the eigenmode spectrum of a model with the fiducial
groove function modified according to the prescription \beqn f_{\rm
  groove}(x) \rightarrow \min \left\{ 1, f_{\rm groove}(x) \right\}
\neqn so that no stars are ever added but only removed. This removes
the ridges at more radial orbits but leaves the grooves at the loci of
near-circular orbits in place. This model has two growing eigenmodes,
at $\omega=0.593 + 0.013 {\mathi}$ and $\omega=1.663 + 0.003
{\mathi}$. These are virtually the same as the eigenmodes of the
fiducial grooved Mestel disc. Likewise, we analysed a model in which
the depth of the three grooves of the fiducial model is kept constant
along lines of constant Jacobi integral. In other words:~the first
groove depresses the DF by 57~\% and the second and third groove by
5~\% everywhere along the groove in question. The stellar mass of this
model is a bit smaller than in the fiducial model but this has no
apparent consequences. Indeed, this model's mode spectrum is almost
identical to that of the fiducial model.

Putting these results together, we are led to the conclusion that it
is the depopulation of the near-circular orbits and {\em not} the
overpopulation of the radial orbits that is crucial to the existence
of the growing eigenmodes of the grooved Mestel disc. This
corroborates the results from \cite{Sellwood91}, who use an
approximate analytical computation to demonstrate that grooves are
destabilizing while ridges are not.

\subsection{The groove depth}

\begin{table}
	\centering
	\caption{Reprise of Table \protect{\ref{tab:mode_table}}, with
          the eigenmodes of the half-mass Mestel disc with three
          grooves at $J_\phi=1.2$, $1.55$, and $1.9$ with various
          depths. The depth of the first groove, expressed relative to
          its fiducial depth, is indicated in the first column. The
          depths of the other two grooves are varied
          proportionally. The complex mode frequency is denoted by
          $\omega$. The radii of the main resonances (ILR, CR, OLR)
          are indicated and resonances that (approximately) overlap
          with the position of a groove are printed in boldface. The
          physical nature of each mode (global or groove mode) is
          given in the last column. }
	\label{tab:mode_table_deeper}
\begin{tabular}{c|c|c|c|c|c}\hline
   depth & $\omega$ & ILR & CR & OLR \\ \hline 155~\% & $0.582 + 0.019
   {\mathi}$ & 1.00 & 3.44 & 5.86 & global mode \\ & $1.709 + 0.016
   {\mathi}$ & 0.34 & {\bf 1.17 } & 2.00 & groove mode \\ & $1.078 +
   0.010 {\mathi}$ & 0.54 & {\bf 1.86 } & 3.17 & groove mode\\ &
   $0.471 + 0.001 {\mathi}$ & {\bf 1.24} & 4.25 & 7.25 & global mode\\ \hline
   100~\% & $0.597 + 0.013 {\mathi}$ & 0.98 & 3.35 & 5.72 & global
   mode \\ &$1.662 + 0.004 {\mathi}$ & 0.35 & {\bf 1.18} & 2.01 &
   groove mode \\ &$0.465 + 0.001 {\mathi}$ & {\bf 1.26} & 4.31 & 7.35
   & global mode \\ \hline 80~\% & $0.604 + 0.008 {\mathi}$& 0.97&
   3.31& 5.65 & global mode\\ \hline 60~\% & $0.611+0.002{\mathi}$ &
   0.96 & 3.27 & 5.59 & global mode \\ \cline{1-6}
\end{tabular}
\end{table}

We varied the depths of the grooves, keeping their respective depths
in proportion. This leads to the eigenmode spectra listed in Table
\ref{tab:mode_table_deeper}. Clearly, the deeper the grooves, the more
rapidly the mode with pattern speed $m\Omega_{\mathrm{p}} \approx 0.6$
grows. This confirms the finding by
\citet{2012ApJ...751...44S,2014ApJ...785..137S} that new instabilities
only start to grow when the phase-space grooves are sufficiently deep.

Moreover, new eigenmodes appear as the grooves are deepened. For
instance, the mode with frequency $\omega = 1.078 + 0.010 {\mathi}$,
which appears in the spectrum of the model with a first groove that is
1.55 times as deep as in the fiducial model, is a groove mode with its
CR at the position of the third groove, at angular momentum
$J_\phi=1.9$. This corroborates the results from \cite{Sellwood91},
who find that a groove of a given width needs to exceed a critical
depth before it can create a groove mode. Using approximate analytical
calculations, these authors predict the critical depth to be
proportional to the groove's width, which is in rough agreement with
the required groove depths we observe here.

\subsection{The position of the grooves}

\begin{figure}
\includegraphics[trim=10 15 -10 0, clip, width=0.495\textwidth]{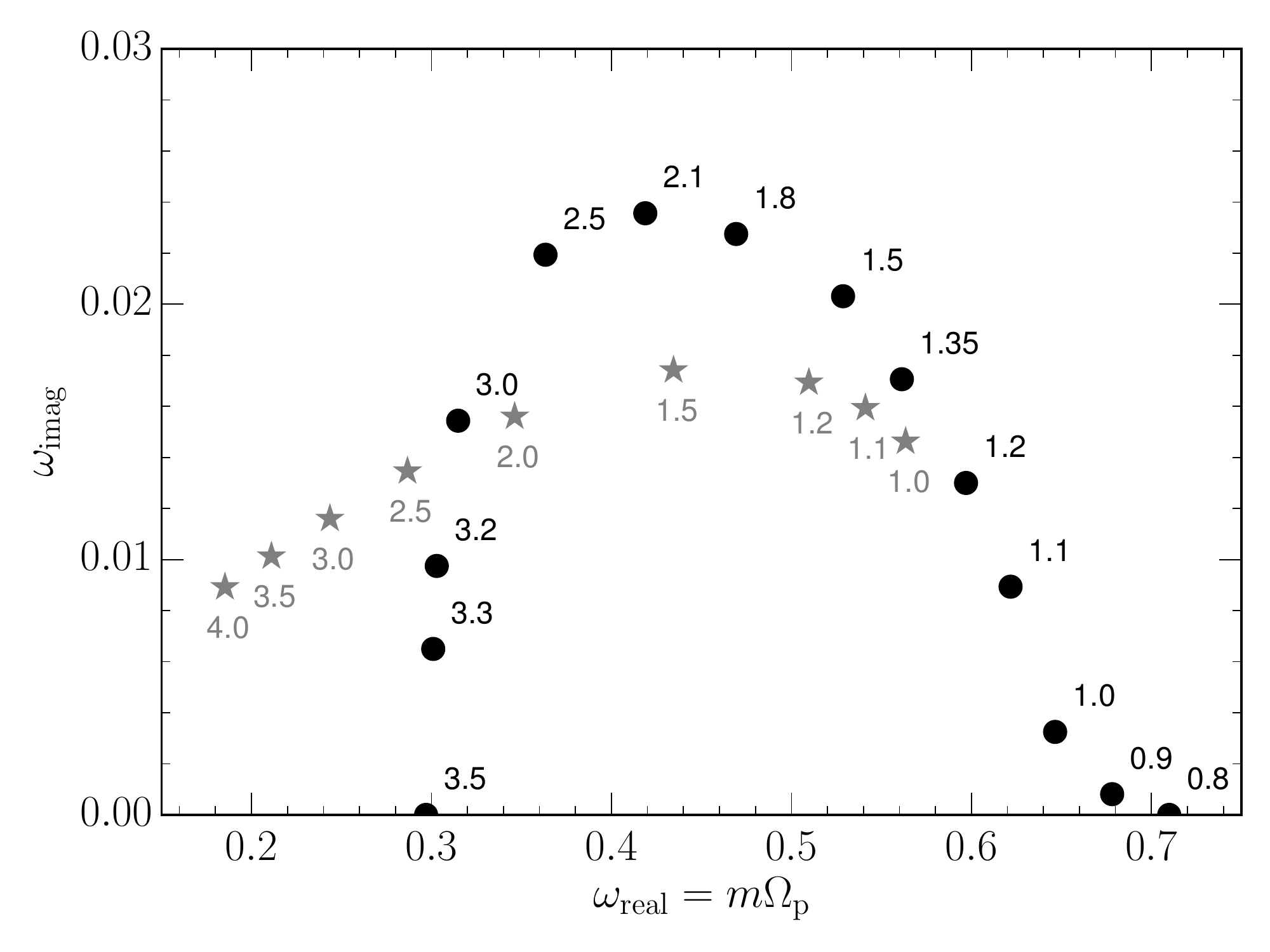}
\caption{Position of the dominant $m=2$ eigenmode in the complex
  frequency plane of the half-mass Mestel disc as a function of groove
  position, computed using the full linear theory (black bullets) and
  using the approximate calculation from paragraph \ref{anarg} (grey
  stars). The angular momentum of the first groove takes on values
  between $J_\phi=0.8$ and $J_\phi=3.5$, as indicated next to each
  data point. The ratio of the angular momentum of the second and the
  third groove to that of the first groove is kept constant, causing
  the grooves to jointly sweep through phase space. The fiducial
  grooved Mestel disc corresponds to $J_\phi=1.2$.
 \label{fig:grooveposition}}
\end{figure}

An investigation of the role of the positions of the grooves is not
possible with the techniques of \citet{2012ApJ...751...44S} and
\citet{2015A&A...584A.129F}, where the grooves grow due to stellar
dynamics and their position is determined by the physics of the
problem. Here, on the contrary, we have full control over the
positions of the grooves. We simply change the angular momentum of the
circular orbits depleted by the first groove between 0.8 and 3.5 while
maintaining a constant pattern speed ratio between the three grooves
and keeping all other groove properties constant. This causes the
grooves to jointly sweep accross the phase-space region with the
highest stellar density.

The result is plotted in Figure \ref{fig:grooveposition}. Here, the
black bullets indicate the frequency of the dominant global eigenmode in
the complex frequency plane for different positions of the
grooves. Each data point is labeled with the angular momentum $J_\phi$
of the circular orbits depleted by the first groove as it sweeps
through phase space. Growing eigenmodes are triggered only if the
grooves fall in a ``responsive'' region of phase space (see also
\citet{dv16}). In this case, this means a groove carving into the most
densely populated region of phase space:~that of the circular orbits
with radii between 1.5 and 2.5. A groove falling outside this region
has a much less marked effect, as can be expected from removing stars
from sparsely populated phase-space surroundings.

The grey stars in Figure \ref{fig:grooveposition} show the analytical
estimate for the frequency of the dominant global eigenmode as a
function of groove position, using the analytical formalism detailed
in paragraph \ref{anarg}. Each star is labeled with the angular
momentum $J_\phi$ of the circular orbits depleted by the first
groove. This estimate, derived from an analytical calculation based on
the assumption that we are dealing with a cavity mode between the
groove radius and the forbidden region around corotation, is clearly
in rough agreement with the full linear theory result. Given the
approximations underlying this calculation, this level of agreement is
actually surprisingly good and can be taken as strong evidence for our
interpretation of the dominant global eigenmode as a cavity mode.

\subsection{Gravitational softening}

\begin{table}
	\centering
	\caption{The eigenmodes of the unsoftened half-mass Mestel
          disc with three grooves at $J_\phi=1.2$, $1.55$, and
          $1.9$. The complex mode frequency is denoted by
          $\omega$. The radii of the main resonances (ILR, CR, OLR)
          are indicated and resonances that (approximately) overlap
          with the position of a groove are printed in boldface. The
          physical nature of each mode (global or groove mode) is
          given in the last column. }
	\label{tab:nosoft_mode_table}
\begin{tabular}{c|c|c|c|c}\hline
  $\omega$ & ILR & CR & OLR & type \\ \hline
  $ 1.726 0+ 0.068 {\mathi}$ & 0.34 & {\bf 1.16} & 1.98 & groove mode  \\
  $ 0.674 0+ 0.040 {\mathi}$ & 0.87 & 2.97 & 5.06 & global mode  \\
  $ 0.547 0+ 0.011 {\mathi}$ & 1.07 & 3.65 & 6.24 & global mode  \\
  $ 1.065 0+ 0.004 {\mathi}$ & 0.55 & {\bf 1.88} & 3.21 & groove mode  \\
  $ 0.542 0+ 0.003 {\mathi}$ & 1.08 & 3.69 & 6.29 & global mode  \\
  $ 0.420 0+ 0.001 {\mathi}$ & 1.39 & 4.76 & 8.13 & global mode  \\
  \cline{1-5}
\end{tabular}
\end{table}

Gravitational softening with a Plummer kernel with softening length
$\varepsilon = 1/8$, as employed by \citet{2012ApJ...751...44S}, has a
very strong impact on the eigenmode spectrum of the grooved Mestel
disc. In Table \ref{tab:nosoft_mode_table}, we list the dominant
eigenmodes of the unsoftened grooved Mestel disc model. Without
gravitational softening, the groove modes grow much faster than in the
softened model. For instance, the groove mode caused by the first
groove is now by far the fastest growing mode. The global modes around
$m \Omega_{\mathrm p} \approx 0.45$ and $m \Omega_{\mathrm p} \approx
0.6$ appear to be relatively unaffected by softening but they are now
surpassed in growth rate by the groove modes. Moreover, new eigenmodes
appear in the spectrum of the unsoftened model:~the very rapidly
growing mode at $\omega=0.674 0+ 0.040 {\mathi}$ and the slowly
growing mode at $\omega= 0.542 0+ 0.003 {\mathi}$.

\section{Conclusions} \label{sec:conc}

We have studied the linear stability of a half-mass Mestel disc, with
a distribution function modified with phase-space grooves like the
ones found in the works of \citet{2012ApJ...751...44S},
\citet{2015A&A...584A.129F}, and \citet{2017fouvry.book}. The
ungrooved model is linearly stable so that any modes found in the
grooved models must somehow be related to the presence of the
grooves. Since we compare our linear stability computations with
$N$-body simulations, we take into account the effects of
gravitational softening \citep{wijzelve}.

In the fiducial grooved model, we mimic the grooves visible in the DF
of the $N$-body simulation 50M of \citet{2012ApJ...751...44S} and show
that this model has $m=2$ eigenmodes with properties very close to the
two-armed spiral patterns visible in its simulated analog. 
We argue that, at least in this case, the dominant new eigenmode is a
cavity mode, living inside a resonance cavity between the groove --
rendered reflective by the removal of stars from near-circular orbits
-- and the forbidden region around corotation
\citep{mark77,2014dyga.book}. Other, more slowly growing new
eigenmodes can be interpreted as groove modes \citep{Sellwood91}.

This confirms the suggestion put forward in
\citet{2012ApJ...751...44S} and \citet{2014ApJ...785..137S} that
grooves can be a potent source of new, rapidly growing
eigenmodes. These grooves are produced by a series of uncorrelated
swing-amplified waves, sourced by the collisional, finite-$N$ nature
of any self-gravitating stellar system. In other words:~it is possible
for an isolated, linearly stable stellar disc, initially without
recourse to any growing eigenmodes that can transport its angular
momentum outwards as desired by the second law of thermodynamics
\citep{lyn72,zhang96}, to spontaneously become linearly unstable via
the formation of grooves in phase space through finite-$N$
dynamics. The simple model of the half-mass Mestel disc with inner and
outer cut-outs serves as a template for more realistic galaxy models,
equipped with more plausible bulge and halo components. The idea is
that the dynamical processes leading to the vigorously growing spiral
modes in this Mestel disc also explain the spirals in $N$-body
simulations of more complex galaxy models, something which needs to
investigated further.

Ultimately, one hopes that understanding these dynamical processes in
simulated galaxies furthers our understanding of the growth and
maintenance of spiral structure in real galaxies. Real spiral galaxies
evidently contain much more stars than the $\sim 10^6-10^8$ particles
that are routinely used in $N$-body simulations and their stellar
bodies are, therefore, much smoother than those of their simulated
counterparts. As shown by \citet{2012ApJ...751...44S} and
\citet{2015A&A...584A.129F}, the time at which the disc's collisional
dynamics has carved sufficiently deep phase-space grooves and the new
eigenmodes start to manifest themselves grows with the number $N$ of
particles so one can wonder whether the mechanism discussed here
actually applies to real galaxies. However, whereas the stellar bodies
of real galaxies are perhaps much smoother than those of simulated
galaxies, they possess other potent sources of gravitational
``noise'', such as giant molecular clouds or globular clusters moving
in or through the stellar disc \citep{ovh13}, orbiting satellite dwarf
galaxies \citep{2008ApJ...689..184M,2018MNRAS.481..286L}, and
dark-matter sub-halos
\citep{2008ASPC..396..321D,2018MNRAS.480.4244C,2018MNRAS.478.1576H},
that could play the same role as the particle noise present in
simulations. Moreover, stellar discs are dynamically cold systems so
that through self-gravity any perturbation is dressed by a
polarisation cloud, which further hastens the disc's evolution.

\section*{Acknowledgements}

We dedicate this paper to the memory of Donald Lynden-Bell
(1935-2018). The study of spiral patterns in disc galaxies is but one
of many fields of astronomy that he enriched with inspiring and
crucial contributions.

We wish to thank Prof. Jerry Sellwood for kindly providing us with
additional information about the $N$-body simulations presented in
\citet{2012ApJ...751...44S}. SDR acknowledges finacial support from
the European Union's Horizon 2020 research and innovation programme
under the Marie Sk{\l}odowska-Curie grant agreement No 721463 to the
SUNDIAL ITN network. JBF acknowledges support from Program number
HST-HF2-51374 which was provided by NASA through a grant from the
Space Telescope Science Institute, which is operated by the
Association of Universities for Research in Astronomy, Incorporated,
under NASA contract NAS5-26555. This research is carried out in part
within the context of Spin(e) (ANR-13-BS05-0005,
http://cosmicorigin.org).

\bibliographystyle{mn2e} 
\bibliography{manuscript}

\appendix

\section{Gravitationally softened linear stability analysis} \label{pystab}

In short:~an axially symmetric disc galaxy model is characterized by a
distribution function $F_0(E,J_\phi)$, with $E$ the specific binding
energy and $J_\phi$ the specific angular momentum of a stellar orbit,
and a (positive) gravitational binding potential $V_0(r)$. {\sc
  pyStab} retrieves those complex frequencies $\omega$ for which a
spiral-shaped perturbation of the form \beqn V_{\rm pert}(r,\phi,t) =
V_{\rm pert}(r)\mathrm{e}^{\mathrm{i}( m\phi-\omega t)} \neqn
constitutes an eigenmode with infinitesimal amplitude. Here,
$(r,\phi)$ are polar coordinates in the stellar disc, $V_{\rm
  pert}(r)$ is a complex function quantifying the mode's amplitude and
phase, $m$ is its multiplicity, $\Omega_{\mathrm{p}} = \Re\{\omega\}/m$ its
pattern speed, and $\Im\{\omega\}$ its growth rate. A general
perturbing potential can always be expanded in such spirals and, for
infinitesimal amplitudes, these can be studied independently from each
other.

In essence, {\sc pyStab} solves the first-order collisionless
Boltzmann equation to find the response distribution function $f_{\rm
  resp}(r,\phi, v_r, v_\phi,t)$ produced by a given perturbation
$V_{\rm pert}(r,\phi,t)$. This response distribution function
generates a response density given by \beqn \Sigma_{\rm
  resp}(r,\phi,t) = \int f_{\rm resp}(r,\phi, v_r, v_\phi,t)
\mathrm{d}v_r \mathrm{d}v_\phi, \neqn which, in turn, corresponds to a
response gravitational potential given by
\begin{equation}
	\label{Vprp}
	V_{\rm resp}(\vec{r}) = G \int\Sigma_{\rm
          resp}(\vec{r}')\,\psi(|\vec{r}-\vec{r}'|)\,\mathrm{d}^2\!\vec{r}',
\end{equation}
with $\psi$ the inter-particle interaction potential. For Newtonian
gravity, one evidently has \beqn \psi(|\vec{r}-\vec{r}'|) =
\frac{1}{|\vec{r}-\vec{r}'|}, \neqn while Plummer softened interactions
are quantified by the choice \beqn \psi(|\vec{r}-\vec{r}'|) =
\frac{1}{\sqrt{\varepsilon^2 + |\vec{r}-\vec{r}'|^2}}, \label{plsoft} \neqn with
$\varepsilon$ the softening length.

Since we want to validate our approach by comparing particular results
with the numerical simulations presented in
\citet{2012ApJ...751...44S}, we mimic the strategies employed in that
work, in particular the softening method. There, the initial condition
is generated by sampling stellar particles from the distribution
function $F_0(E,J)$ evaluated using the mean Newtonian gravitational
potential $V_0(r)$, independent of the gravitational softening that is
employed to evolve the particles through time. Moreover, the axially
symmetric force field of the base state is evaluated correctly,
i.e. without softening, while the non-axisymmetric force field of the
growing waves is softened. Therefore, in our search for unstable
models, we only implement Plummer-type softening in the response
potential $V_{\rm resp}(r,\theta,t)$, but not in the axially symmetric
base state potential $V_0(r)$. Using this strategy, equation (\ref{Vprp})
is the only place where the softened gravitational interaction enters
the computation of the modes. This approach to including gravitational
softening in linear stability analysis is detailed in
\citet{wijzelve}.

Eigenmodes are identified by the fact that \beqn V_{\rm
  pert}(r,\phi,t) \equiv V_{\rm resp}(r,\phi,t), \neqn and {\sc pyStab}
employs a matrix method \citep{kalnajs77,b9} to find them. To do so,
the perturbing potential $V_{\rm pert}$ is expanded in a basis of
potentials, $V_\ell$. The response to each basis potential, denoted by
$V_{\ell,\rm resp}$, can likewise be expanded in this basis as \beqn
V_{\ell,\rm resp} = \sum_{ k} {\mathcal C}_{\ell k} V_{ k}. \label{Cdef} \neqn If
the perturbation is an eigenmode, then the $\omega$-dependent
${\mathcal C}$ matrix can be shown to possess a unity eigenvalue
\citep{b9}. This feature is exploited by {\sc pyStab} to identify the
eigenmodes. Since it is assumed that the mode amplitude is zero at
time $t=-\infty$, we only consider growing modes, i.e. with
frequencies with a positive imaginary part.

The formalism contains a number of technical parameters, such as the
number of orbits on which phase space is sampled (here we use $n_{\rm
  orbit}(n_{\rm orbit}+1)/2$ orbits with $n_{\rm orbit}=600$ in the
allowed triangle of turning points -- or pericentre/apocentre --
space), the number $n_{\rm Fourier}$ of Fourier components in which
the periodic part of the perturbing potential is expanded (here we use
$n_{\rm Fourier}=80$), the number of potential-density pairs (PDPs)
that is used for the expansion of the radial part of the perturbing
potential and density (we typically use 40 PDPs), and the shape and
extent of the PDP density basis functions. Here, we use PDP densities
$\Sigma_\ell$ with compact radial support. They cover the relevant
part of the stellar disc and are evenly spaced on a logarithmic scale
so the resolution is highest in the inner regions of the disc. Their
radial widths are automatically chosen such that consecutive basis
functions are unresolved and can be used to represent any smooth
radial function. The corresponding PDP potentials are computed
numerically using equation (\ref{Vprp}).

\label{lastpage}

\end{document}